# Revealing the role of molecular rigidity on the fragility evolution of glass-forming liquids

C. Yildirim[1,2], J.-Y. Raty[2], M. Micoulaut[1]

[1] Paris Sorbonne Universités – UPMC, Laboratoire de Physique Théorique de la Matière Condensée, Paris Sorbonne Universités – UPMC Boite 121, 4, Place Jussieu, 75252 Paris Cedex 05, France

[2] Physique de la Matière Condensée, B5, Université de Liège, B4000 Sart-Tilman, Belgium

**Abstract:** If quenched fast enough, a liquid is able to avoid crystallization and will remain in a metastable supercooled state down to the glass transition, with an important increase in viscosity upon further cooling. There are important differences in the way liquids relax as they approach the glass transition, rapid or slow variation in dynamic quantities under moderate temperature changes, and a simple means to quantify such variations is provided by the concept of "fragility". Here, we report molecular dynamics simulations of a typical network-forming glass, Ge-Se, and find that the relaxation behaviour of the supercooled liquid is strongly correlated to the variation of rigidity with temperature and the spatial distribution of the corresponding topological constraints which, ultimately connect to fragility minima. This permits extending the fragility concept to aspects of topology/rigidity, and to the degree of homogeneity of the atomic-sale interactions for a variety of structural glasses.

1. **Introduction**

Once the melting point has been bypassed, both viscosity ($\eta$) and structural relaxation time ($\tau_\alpha$) of a supercooled liquid increase enormously as the temperature is lowered. Fragility (M) of glass forming melts has been introduced to quantify such temperature dependences, and a convenient way of obtaining M [1] is to estimate the slope of $log_{10}(\eta)$ with temperature, near the glass transition temperature $T_g$ where, by definition, one has $\eta=10^{12}$ Pa.s, and $\tau_\alpha$~100s.

$$M = \left[ \frac{\partial log_{10}\tau_\alpha}{\partial \left(\frac{T_g}{T}\right)} \right]_{T=T_g} \quad (1)$$

This dimensionless slope M, examined for a broad range of glass forming liquids, has been found to vary between a value of 214 [3] to 14.8 [4], the latter low value being associated with so-called "*strong*" glass formers, whereas liquids with a high M are termed as "fragile". When represented on an appropriate plot (i.e. $log_{10}(\tau_\alpha)$ versus $T_g/T$), strong glass formers (e.g. silica) follow an Arrhenius behaviour of the form $exp[E_A/k_BT]$, while fragile ones will exhibit an important non-linear behaviour and a dramatic increase of $\eta$ (or $\tau_\alpha$) over a limited temperature interval once the supercooled liquid approaches $T_g$. A certain number of correlations between the fragility index M and materials properties have been emphasized, and relationships with $T_g$ [3], structural features [5,6] and elastic properties [7], have been established. These contributions simply reveal that various aspects of physical properties, structure, and interactions influence the glassy relaxation and the fragility.

Here we show by using molecular dynamics (MD) simulations that a typical network-forming system with changing composition displays a certain number of relaxation dynamics anomalies that can be correlated with the strong or fragile nature of the corresponding supercooled liquids. Relaxation properties in network glasses has been quite extensively studied, as in e.g. silica [8]. We adopt here, however, a slightly different approach, albeit in line such previous work [8]. Off-stoichiometric glass-formers, e.g. $Ge_xSe_{100-x}$ with $GeSe_2$ (*x*=33%) being isochemical to silica, afford, indeed, a unique venue for exploring the dynamics, and understanding ultimately the physical origin of fragility and relaxation in network glasses. First, such studies permit to tune the composition in a continuous fashion, and to compare compositional trends in physical properties, leading to the identification of new correlations. Second, by focusing on a combination of several methods, molecular simulations and Rigidity Theory [9,10], one can accumulate direct probes and statistical time-dependent quantities that matter during the glass transition. Third, the focus on key physical behaviour, i.e. the small temperature evolution of network rigidity for certain compositions is seen to cause the strong character of corresponding glass-forming melts and is driven by the spatially homogeneous distribution of constraints/interactions at select compositions. This offers a new interpretation of the physics driving fragility of network glasses.

It is detected that the compositions displaying an anomalous dynamics merely satisfy the well-known Maxwell stability (isostatic) criterion [11], while also weakly depending on temperature. Given that such isostatic compositions also lead to a homogeneous distribution of mechanical constraints at the atomic scale, we argue that fragility deeply connects to the rigidity properties of the underlying network structure, and that fragile glasses are characterized by a distribution of heterogeneous interactions. These conclusions are successfully compared to available experimental data which support the present findings, while also emphasizing the general character of the results for a variety of structural glasses.

## 2. Results

**2.1 Dynamic and relaxation anomalies.** To establish our conclusions, we performed First Principles Molecular Dynamics (MD) of $Ge_xSe_{1-x}$ liquids and glasses at various Ge content (10≤x≤33%). The modelling scheme [12] reproduces accurately structural data accessed from neutron diffraction [17,18] in glasses and liquids (Supplementary Figs. 1-2 and supplementary note 1). The structural changes undergoing in Ge-Se are well documented [13,14,15], and generic to the family of Group IV chalcogenides, i.e. glassy systems made of two-fold Se chains are progressively cross-linked by the addition of four-fold germanium atoms that transforms the basic chain structure into a highly connected network (**Fig. 1a,b**). The addition of such higher coordinated species leads to a global stiffening of the network structure, and by using the notion of Maxwell rigidity [11] and an enumeration of constraints $n_c$ [9,10], a flexible to rigid transition has been identified at the composition $x_c$=20%, confirmed experimentally at room temperature [15], and identified with the mean-field Maxwell isostatic stability criterion $n_c$=3.

The key findings of our study are shown in **Fig. 1c,d**. Different quantities encoding the dynamics and relaxation behavior of the Ge-Se melts (at 1050 K allowing for a validation of the structural models) have been calculated. From the mean-square displacement $<r_i^2(t)>$ one obtains the diffusion constants in the long time limit using the Einstein relation $D=<r_i^2(t)>/6t$. A global decrease of the Ge and Se diffusivity is obtained with increasing Ge content, in line with the stiffening of the network structure that progressively hinders atomic motion. However, a striking feature is the presence of a narrow interval in composition (typically 18-22% Ge) for which $D_{Ge}$ displays an even more marked decrease, about a factor 2-3 less with respect to what would be expected from a smooth decrease of $D_{Ge}$ (broken red line). Further characterization of the dynamics and detecting the anomalous behavior of the Ge diffusivity is provided by the self part of the Van Hove correlation function $4\pi r^2 G(r,t)$ which represents the probability density of finding an atom at time t knowing that this atom was at the origin (r=0) at time t=0. In fact, a minimum for the corresponding jump probability is found (**inset of Fig. 1c**) revealing that the atomic motion will be substantially reduced for this particular interval in composition.

**2.2 Topological constraints.** A central result is the detection of the crucial role played by rigidity in the anomalous relaxation dynamics of **Fig. 1**, and the indication of a fragility minimum. Within temperature dependent Rigidity Theory [16], the assumption of a Adam-Gibbs form for the relaxation time $\ln \tau \propto 1/TS_c$, and the contribution of the number of topological degrees of freedom $f(x,T)=3-n_c(x,T)$ to the configurational entropy [17] $S_c(x,T)$ leads to the prediction of the fragility index M :

$$M = M_0 \left[ 1 + \frac{\partial \ln f(T,x)}{\partial \ln T} \right]_{T=T_g(x)} \qquad (2)$$

which depends only on the scaling of $f(x,T)$ with temperature. The validity and predictive power of equation (2) has been checked on a certain number of glasses by building structural models, establishing their constraint count, and comparing with a composition dependence of fragility measurements such as for borates [18] or phosphates [19]. MD based constraint counting algorithms (Supplementary note 2) are used [19] to yield $n_c(x,T)$ from Ge/Se bond-stretching (BS) and bond-bending (BB) motions, and they show that i) for the 300 K glasses, $n_c$ follows exactly the mean-field estimate $n_c=2+5x$ (**Fig. 2a**) [9,10], leading to an isostatic condition ($n_c=3$) for ~20% Ge that agrees with the experimentally measured threshold value [20], and, importantly, ii) the calculated $n_c(x,T)$ displays a minimum change with temperature for 20-22% Ge. The obtained minimum change is robust (see supplementary note 3). Following equ. (2), this leads to a minimum for the liquid fragility with composition given the weaker variation of the term $dn_c/dlnT$ at x~22%, and coincides with $n_c=3$.

**2.3. Viscosity anomaly.** Independent support for this conclusion is given by a Stokes-Einstein estimate of the liquid viscosity. Given the system size (250 atoms), and First Principles MD timescale (ps), a direct calculation of the Green-Kubo viscosities is out of reach. However, the validity of the Stokes-Einstein $\eta=k_BT/6\pi rD$ where $2r=\rho^{-1/3}$ involves the liquid densities $\rho$, can been established from calculated diffusivities for the supercooled state, similarly to a previous application [20], and this provides a validation of the dynamic behaviour of the models (see supplementary note 4). Next, one can derive from the calculated diffusivities $D_{Ge}(x,T)$ (**Fig. 1a**) corresponding viscosities using the Stokes-Einstein relationship. Results using such calculated viscosities are shown in **Fig. 3**, and a direct comparison in an Angell plot represents calculated and measured log10 $\eta$ as a function of $T_g/T$. Note that we have checked that the Stokes-Einstein remains valid for the considered range of temperatures. The agreement is found to be very good for the two compositions for which a direct comparison is possible ($Ge_{20}Se_{80}$ and $Ge_{33}Se_{67}$). Having validated the approach, we then use it to extract the viscosity $\eta(x,1050K)$ along the 1050 K isotherm (**Fig. 1d**). A maximum for $\eta$ is obtained that correlates close to the isostatic composition (**Fig. 1d**), with the maximum of the structural relaxation time $\tau_\alpha$ calculated from the $\alpha$-relaxation regime of the intermediate scattering function (Supplementary note 5 and supplementary Fig. 5). This provides an additional indication that strong glass-forming liquids with a minimum fragility are obtained for the 22% composition.

3. Discussion

When the details of the constraint contributions are investigated, and their spatial distribution analysed (**Fig. 2b**), additional important features can be linked to the weak variation of $n_c$ with temperature. First, for the composition at which the weakest change in $n_c$ is found (22%), the BB constraints are nearly homogeneously distributed in the structure given that $n_c^{BB}$ maximizes to its nearly low temperature value ($n_c^{BB}$=5) which reduce possible fluctuations (**Fig. 2c**). This contrasts significantly with all the other compositions (e.g. 14.3 and 33%, **Fig. 2b**) which display a heterogeneous distribution of constraints broken by thermal activation (**Fig. 2d**). The variance of $n_c^{BB}$ is also a measure of the spread in the spatial distribution of atomic interactions leading to rigid constraints. For the compositions in the region 20-23%, one now acknowledges that homogeneity of stress (i.e. interactions) will not only induce a cascade of dynamic anomalies as detected from the diffusivity minima and viscosity/relaxation time maxima located in the same range of compositions (**Fig. 1**) but also represents a more stable state at the nanoscale. As a consequence, thermal changes must lead to minor changes of $n_c$ with temperature, and ultimately will minimize fragility (equ. (2)). In this particular region of anomalous behavior, one furthermore detects weak effects on structure with chemical composition given that both the pair correlation function g(r) in real space and the structure factor in reciprocal space exhibit small differences with Ge content (see supplementary note 1 and supplementary Figs. 1, 2). For instance, the structural properties of the 22% composition which appears to be so particular in terms of dynamic properties (**Fig. 1**), are found to be nearly identical to the 20% one, although their relaxation time differs by a factor 2 (**Fig. 1b**) and their rigidity also varies substantially within a tiny compositional interval. This, ultimately, emphasizes the predominance of rigidity and its spatial distribution on the liquid relaxation properties, over aspects of structure.

This physical picture revealing the central role played by network rigidity is consistent with conclusions drawn from a compilation of fragility data on network glasses including chalcogenides and oxides for which the location of the isostatic composition is determined from calorimetric measurements [4,26-29]. These liquid fragilities are represented in **Fig. 4**, and show that once the composition is rescaled with respect to the centroid $x_c$ of the isostatic composition interval (e.g. $x_c$=22.5% in $Ge_xSe_{100-x}$ [4]), all data exhibit a minimum in M at $x \sim x_c$. This not only emphasizes that rigidity affects the fragility evolution of melts, a qualitative correlation that has been reported in the literature for quite some time ([4] and reference therein), but from **Fig. 2b**, one also realizes that the spatial distribution of atomic scale interactions/constraints is a key feature for the understanding of transport properties during the glass transition. This conclusion can be only drawn from an detailed inspection of constraints, accessed from MD on individual atoms. Although further detailed analysis on the constraint behavior of e.g. organic glass formers is necessary, these results may indicate that liquids with non-directional bonding and, therefore, with a more probable heterogeneous distribution of interactions will lead to a much more fragile behavior.

**Methods**

**First Principles molecular dynamics simulations.** Ge-Se liquids and glasses have been investigated using Car-Parrinello molecular dynamics (CPMD) simulations. The system contained 250 atoms, and up to 9 compositions $Ge_xSe_{100-x}$ have been simulated in NVT ensemble with cubic cells of sizes allowing to recover the experimental density [30,31] of corresponding liquids and glasses (i.e. 19.97 Å for $GeSe_9$).

Density functional theory (DFT) has been used to describe the electronic structure that evolved self-consistently with time. We have adopted a generalized gradient approach (GGA) using the exchange energy obtained by Becke [32] and the correlation energy according to Lee, Yang and Parr (LYP) [33]. The BYLP approach was used due to its account on valence electron localization effects, and an increased agreement with experiments on structure as revealed by a previous study [12,34]. Valence electrons have been treated explicitly, in conjunction with norm conserving pseudopotentials of the Trouiller-Martins type to account for core-valence interactions. The wave functions have been expanded at the Γ point of the supercell on a plane-wave basis set having an energy cut-off $E_c$ = 20 Ry which is a standard value for the investigation of chalcogenides [35]. A fictitious electron mass of 200 a.u. was used in the first-principles molecular dynamics (FPMD) approach. The time step for integrating the equations of motion was set to $\Delta t$ = 0.10 fs. The temperature control was achieved for both ionic and electronic degrees of freedom using Nosé-Hoover thermostats. The initial coordinates of 250 atoms have been constructed using the atomic positions in a GeSe crystal. To achieve the correct compositions, Ge atoms were then replaced with Se atoms in an appropriate fashion, depending on the target composition.

We carried out simulations at T = 2000 K for a period of 22ps in order to lose the memory of initial configuration, and then investigated a certain number of isotherms (for 25-30 ps each): 2000 K, 1600 K, 1373 K, 1200 K, 1050 K, 800 K, and 300 K. The first 2 ps at all temperatures have been discarded. The 300 K trajectories result from three independent quenches with starting temperature 1050 K.

Certain compositions have been particularly checked, e.g. the 22% and 23%, and a series of five trajectories at 1050 K obtained from five independent high temperature liquid starting configurations at 2000 K have been analysed. The statistical average calculated quantities (e.g. $n_c$, $n_c^{BB}$, $D_{Se}$, etc.) have been found to be consistent (see supplementary note 3). Similarly, effects of size (480 atoms and the present 250 atoms) and excursion in density have been tested and have confirmed that the constraint count does not depend on these different simulation conditions. For instance, for the 23% composition, it has been found that $n_c$=2.84 for N=480, a value that is very close to the value found for N=250 ($n_c$=2.79).

**Acknowledgements** – This work is supported by Agence Nationale de la Recherche, and National Science Foundation (International Materials Institute). MM acknowledges support from a Fulbright Fellowship. C.Y. acknowledges IDS FunMat for the financial support for his PhD project (Project 2013-05-EM). Consortium des Equipements de Calcul Intensif (CECI, funded by F.N.R.S) is acknowledged for supercomputing access. We thank D. Skoncz-Traintz for technical support. P. Boolchand is acknowledged for stimulating discussions and for sharing his data on glass-forming melts.

**Author contributions**

M.M. conceived the original idea, C.Y. performed the molecular dynamics simulations, C.Y., J.-Y. R. and M.M analysed the simulation results and built the figures. J.-Y. R. and M.M. wrote the paper.

[5] N. A. Mauro et al., A structural signature of liquid fragility, Nat. Comm. **5**, 4616 (2014)

[6] D.L. Sidebottom, T.D. Tran, S.E. Schnell, Building up a weaker network: The effect of intermediate range glass structure on liquid fragility, J. Non-Cryst. Solids **402**, 16-20 (2013).

[7] T. Scopigno et al., Is the Fragility of a Liquid Embedded in the Properties of its Glass ? Science **302**, 849-852 (2003).

[8] M. Bauchy, M. Micoulaut, Densified network glasses and liquids with thermodynamically reversible and structurally adaptive behavior, Nat. Comm. **6**, 6398 (2015).

[9] J.C. Phillips, Topology of covalent non-crystalline solids I: Short-range order in chalcogenide alloys. J. Non-Cryst. Solids **34**, 153-181 (1979).

[10] H. He, M.F. Thorpe, Elastic properties of glasses. Phys. Rev. Lett. **54**, 2107–2110 (1985).

[11] J.C. Maxwell, On the calculation of the equilibrium and stiffness of frames, Philos. Mag. **27**, 294-299 (1864). J.-L. Lagrange, Mécanique Analytique (Paris, 1788).

[12] M. Micoulaut et al., Improved modeling of liquid GeSe$_2$: Impact of the exchange-correlation functional, Phys. Rev. B **79**, 214205 (2009)

[13] P.S. Salmon et al., Topological versus chemical ordering in network glasses at intermediate and extended length scales, Nature **435**, 75 (2005)

[14] P.S. Salmon, Structure of liquids and glasses in the Ge-Se binary system. J. Non-Cryst. Solids **353**, 2959-2974 (2007).

[15] X. Feng et al., Direct Evidence for Stiffness Threshold in Chalcogenide Glasses. Phys. Rev. Lett. **78**, 4422-4426 (1997)

[16] P. K. Gupta, J. C. Mauro, Composition dependence of glass transition temperature and fragility. I. A topological model incorporating temperature-dependent constraints. J. Chem. Phys. **130**, 094503 (2009).

[17] G.G. Naumis, Energy landscape and rigidity, Phys. Rev. E **71**, 026114 (2005)

[18] J.C. Mauro, P.K. Gupta, R.J. Loucks, *Composition dependence of glass transition temperature and fragility II. A topological model of alkali borate liquids*, J. Chem. Phys. 130, 234503 (2009).

[19] C. Hermansen, J.C. Mauro, Y. Yue, *A model for phosphate glass topology considering the modifying ion sub-network*, J. Chem. Phys. 140, 154501 (2014).

[20] M. Micoulaut et al., Understanding phase-change materials from the viewpoint of Maxwell rigidity. Phys. Rev. B **81**, 174206 (2010).
9

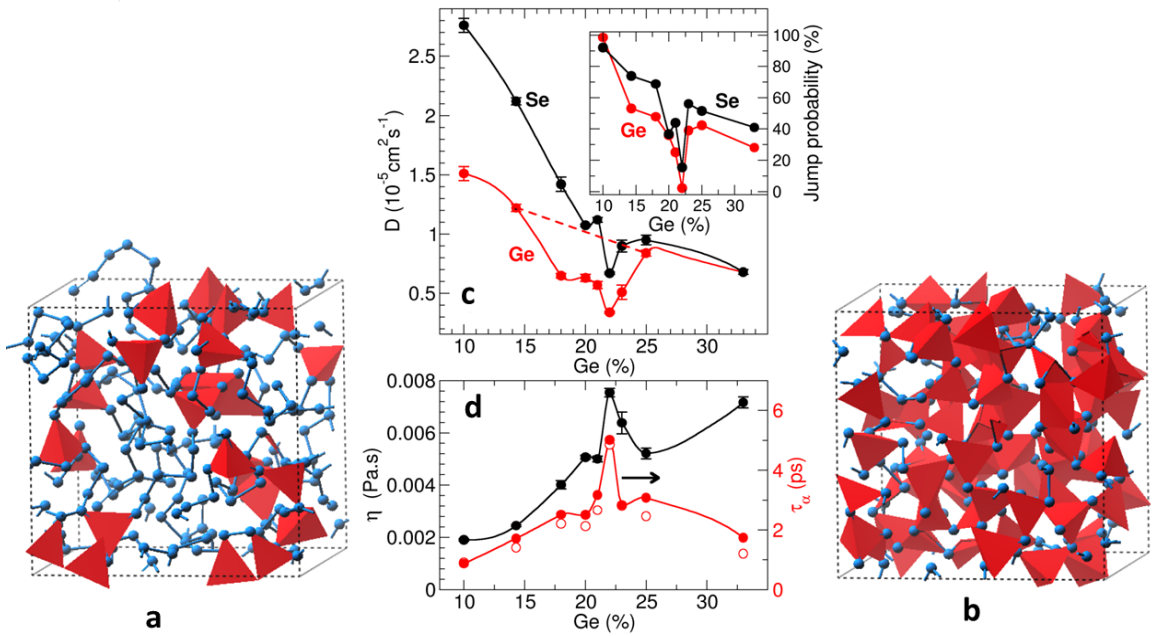

**Figure 1 | Dynamic and relaxation anomalies in Ge-Se glass-forming melts** At low Ge content, the structure of Ge-Se liquids is made of a Se chains with few GeSe$_{4/2}$ tetrahedral cross-links (MD snapshots of Ge$_{10}$Se$_{90}$, **a**), whereas the structure of the stoichiometric GeSe$_2$ (**b**) is made of a network of connected tetrahedra. **c)** Calculated Ge and Se diffusion constants D at 1050 K as a function of Ge content. The inset shows the calculated jump probability that a Ge/Se atom has jumped by a distance r>2.88 Å during t=20 ps. **d)** Calculated Stokes-Einstein liquid viscosities η and Kohlrausch fitted relaxation time $\tau_\alpha$ of the calculated intermediate scattering function F$_s$(k,t) (right axis) at 1050 K as a function of Ge content. The open symbols ($\tau_\alpha$) refer to the particular value F$_s$(k,$\tau_\alpha$)=1/e. Error bars are in most of the cases of the size of the symbols but some are visible.

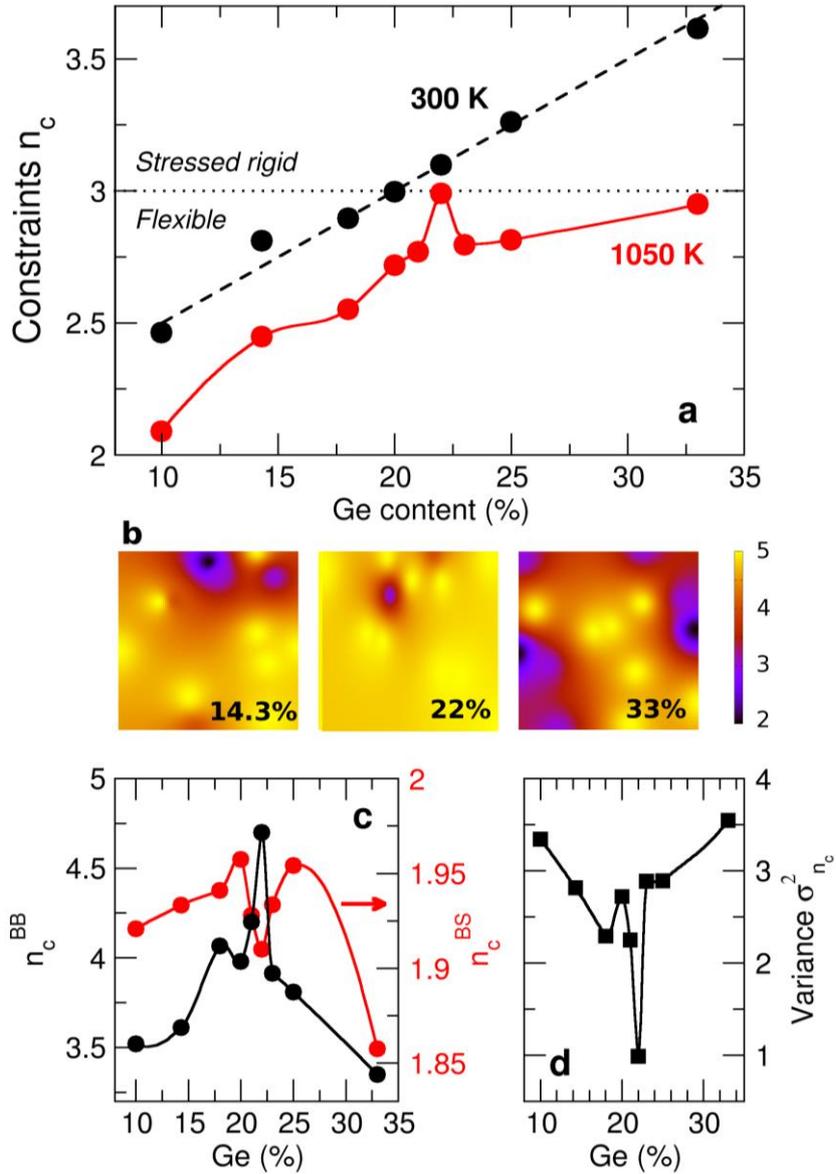

**Figure 2 | Constraint analysis of Ge-Se melts a)** Global MD constraint count of the liquid (1050 K), compared to a count in the glassy state (300 K). The broken line represents the Phillips-Thorpe constraint count $n_c=2+5x$ [12]. **b)** Contourplots of the Ge BB constraint distribution among the structure for selected compositions. They have been generated by focusing on the constraint distribution inside a slab of 3.2±0.2 Å. **c)** Calculated Ge BB (black) and BS (red, right axis) constraint density as a function of Ge content. **d)** Variance $\sigma_{nc}$ of the Ge BB constraint population.

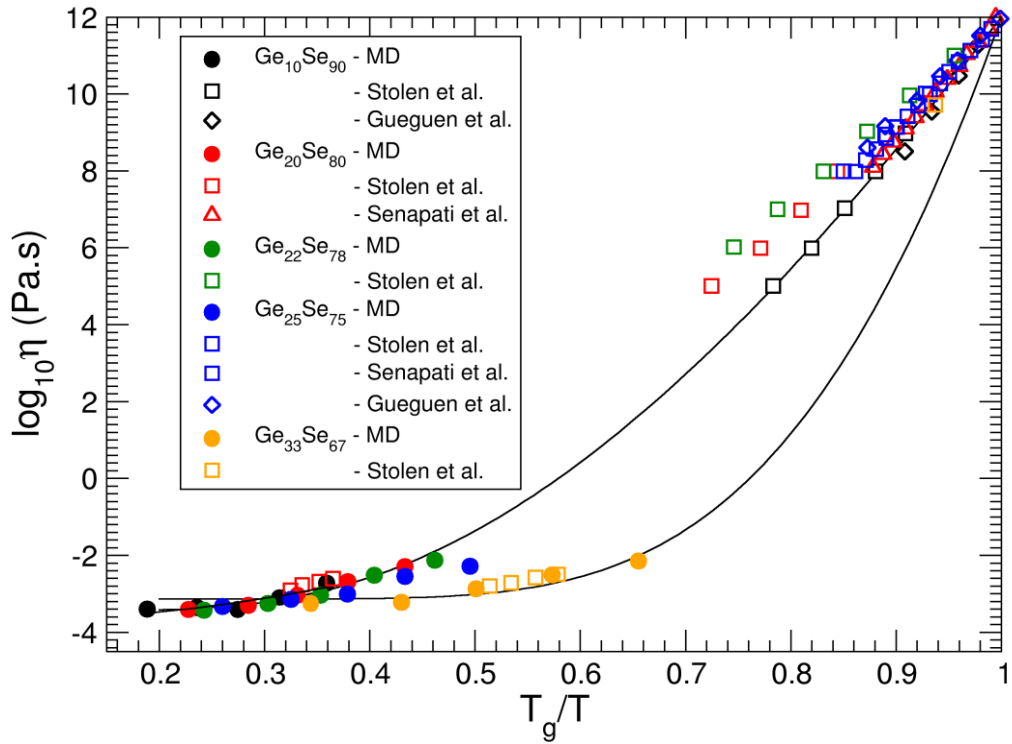

**Figure 3 | Comparison between estimated and measured viscosities** Comparison between the Stokes-Einstein calculated (filled symbols) viscosities at various Ge-Se compositions and experimental measurements of viscosity (open symbols) from Stølen et al. [22], Senapati et al. [23] and Guegen et al. [24]. Note that only Stølen et al. have investigated the high temperature region where numerical data can be directly compared. The solid lines correspond to fits using the Mauro-Yue-Ellison-Gupta-Allen (MYEGA) equation [25] that describes with an increased accuracy the high temperature viscosities as compared to alternative fitting functionals (e.g. Vogel-Fulcher-Tamman, VFT).

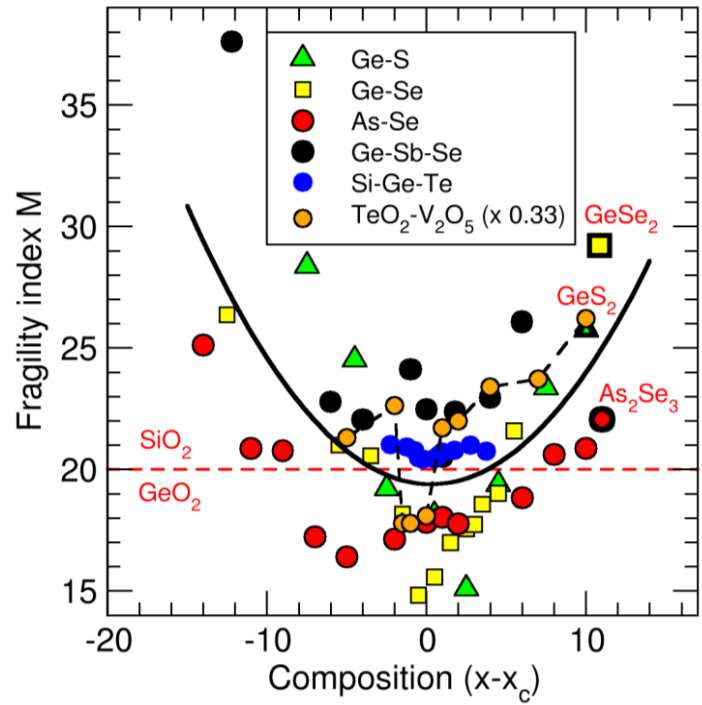

**Figure 4 | A summary of experimentally measured fragilities in glass network-forming melts** Experimental fragilities as a function of a rescaled composition corresponding to the centroid $x_c$ of the corresponding isostatic window (or Boolchand phase): Ge-S [26], Ge-Se [4], As-Se [27], Ge-Sb-Se [28], Si-Ge-Te. All display a minimum at or close to $x=x_c$. Note that the minimum for Ge-Si-Te is clearly visible if represented on an appropriate scale [29]. The trend in composition for $TeO_2$-$V_2O_5$ is highlighted by a thin broken line. The solid line is a global quadratic fit to the whole data sets, and serves only as guide. Stoichiometric compounds are signalled.

# Supplementary information

**Supplementary note 1: Validation of the structural models.**

Prior to the investigation of the dynamics, we have checked that our structural models were realistic enough. In the glass (300 K), a certain number of quantities can be directly compared to results from neutron diffraction [1], including the pair correlation functions *g(r)* (Supplementary Figure 1) and structure factors *S(k)* (Supplementary Figure 2).

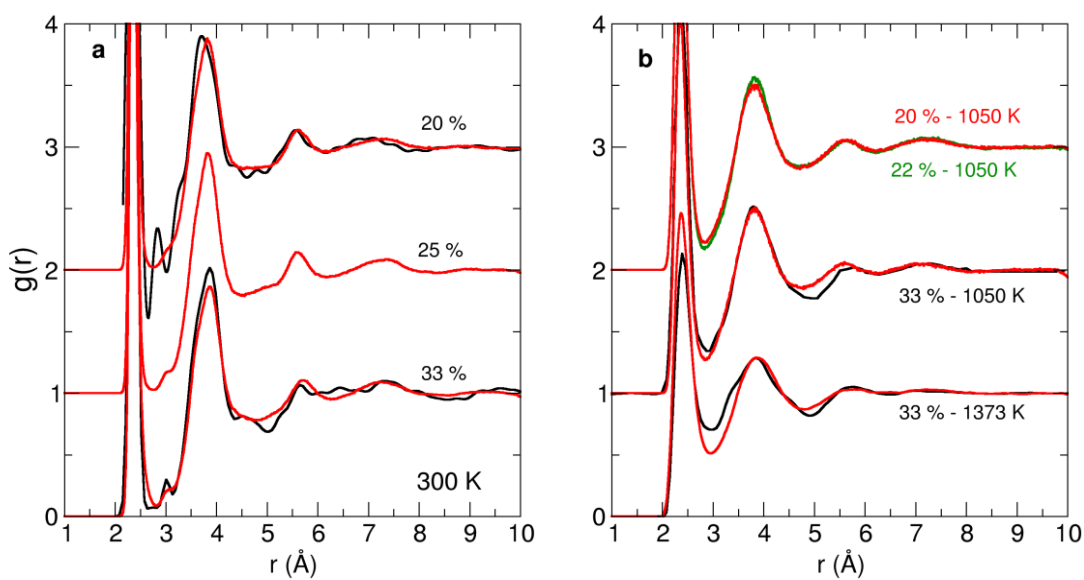

*Supplementary Figure 1: Calculated total pair correlation function g(r) in Ge-Se glasses (a) and melts (b) for different Ge content (red), and compared to available measurements from neutron diffraction (black) [1] at 300 K, 1050 K and 1373 K. The green curve corresponds to the pair correlation function of $Ge_{22}Se_{78}$.*

In real space, the computed total pair distribution function g(r) for the $Ge_xSe_{1-x}$ glasses and liquids compare very convincingly to the experimental counterparts (Supplementary Figure 1) [1,2]. All the features of the pair distribution function are reproduced such as the main peak at 2.36Å, (experimentally [1], one has 2.35 Å) and the secondary peak at 3.85 Å which is due to the Se-Se distance defining the edge of the $GeSe_{4/2}$ tetrahedra. The large intensity of the first peak, which mostly arises from the Ge-Se bond distance, actually overwhelms in Ge-rich glasses other contributions due to Ge-Ge and Se-Se correlations in the structure, and which has been investigated from a detailed analysis of the partial pair correlation functions [1].

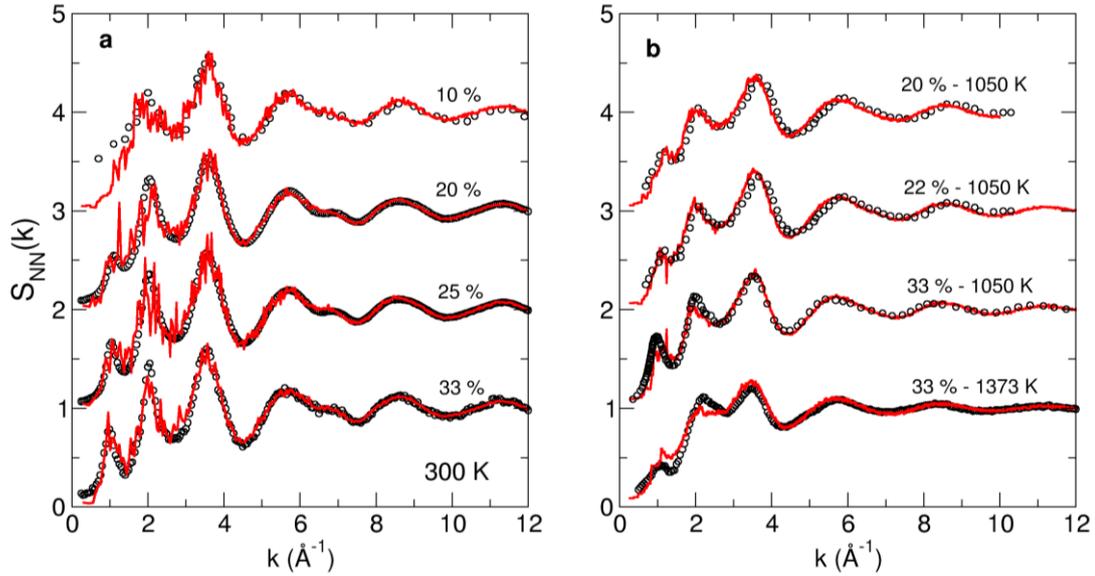

*Supplementary Figure 2: Calculated Bhatia-Thornton structure factor $S_{NN}(k)$ in Ge-Se glasses (a) and melts (b) for different Ge content (red), and compared to available measurements (circles) from neutron diffraction: $Ge_{10}Se_{90}$ [2], $Ge_{20}Se_{80}$ [3], all other compositions/temperatures [1]. The calculated structure factor of $Ge_{22}Se_{78}$ is compared to experimental results of $Ge_{20}Se_{80}$ [3].*

Regarding the reciprocal space, an excellent overall agreement of the calculated $S_{NN}(k)$ with the experimental counterpart is found (Supplementary Figure 2), except for a notable underestimation of the main peaks amplitude at the highest temperatures. All typical peaks of the diffraction pattern are reproduced over the entire range of wave vectors *k*: the first sharp diffraction peak (FSDP) at $k = 1.0$ Å$^{-1}$, the first principal peak at 2.1 Å$^{-1}$, and the second principal peak at 3.6 Å$^{-1}$ The higher wavevector region ($k = 10$–$12$ Å$^{-1}$) is also very well reproduced, including a small shoulder peak at around 6.5 Å$^{-1}$. This agreement is verified for several available compositions in the glassy state (300 K) and for certain liquid temperatures (1050 K and 1373 K). It is to be noticed that small changes in composition (e.g. 1% Ge) do not change substantially the calculated diffraction pattern. For instance, the functions *g(r)* and $S_{NN}(k)$ of $Ge_{21}Se_{79}$ and $Ge_{22}Se_{78}$ (green curve in Supplementary Figures 1 and 2) are found to be very close to the one of $Ge_{20}Se_{80}$. Therefore, the difference in dynamic properties between 20 and 22% Ge (Fig. 1) cannot rely on structural changes.

**Supplementary note 2: MD based constraint counting**

In order to evaluate the number of topological constraints represented in **Fig. 2**, we have calculated the radial and angular excursions between pairs or triplet of atoms, based from the MD atomic configurations at fixed Ge content and temperature. This enumeration is directly inspired by the classical mechanics view of mechanical constraints associating large/small radial or angular motion with the absence/presence of corresponding bond-stretching (BS) and bond-bending (BB) restoring forces. It has been shown [4,5] that BS constraints can be simply enumerated from the coordination number $r_i$ of the atoms, and according to Phillips and Thorpe, this leads to a contributions of $r_i/2$ for the BS constraint, each bond/interaction being shared by two neighbors. To derive angular (BB) constraints, one follows the angular motion around each individual atom $k$ ($k$=Ge,Se) defined by a set of two neighbors. Over the time MD trajectory, the corresponding bond angle distribution $P_k(\theta)$ allows defining a mean (the first moment of $P_k(\theta)$) and a standard deviation $\sigma_k$ (the second moment) that shows a bimodal distribution $f(\sigma_k)$ for the various considered conditions. Atoms subject to a rigid bending interaction contribute to the part of $f(\sigma_k)$ with low $\sigma_k$ ($\sigma_k<\sigma_{min}$) and have a corresponding angle that acts as a rigid BB constraint, inducing network stiffening. The threshold value $\sigma_{min}$ is fixed by $f(\sigma_{min})$ minimum. Averages over the whole simulation box then lead to the mean number of constraints $n_c^{BB}$ per atom that is shown in e.g. **Fig. 2c**.

**Supplementary note 3: cross-checking certain important compositions**

For the compositions in the region 20-25%, we have realized a certain number of additional tests of validation in order to check that the results do not depend on e.g. system size or initial configuration. Five independent configurations have been chosen in the high temperature liquid and cooled to the isotherm of 1050 K over the same time interval (20 ps). All led to similar results in terms of constraints, either total number of interaction or related numbers (BS or BB) with a minimal spread between configurations. Supplementary figure 3 shows the calculated values for such configurations which validate e.g. the data point of 22% represented in Fig. 2a, c and d. In addition, the calculated pair correlation function for the five independent configurations at 1050 K of the 22% system lead to identical results (Supplementary figures 1 and 2), and corresponding g(r) are very close to the 20% system.

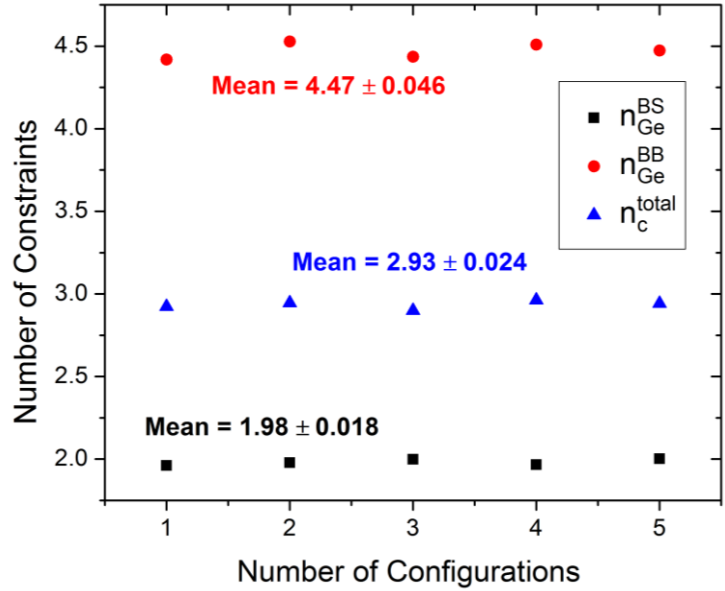

**Supplementary figure 3** Number of constraints $n_c^{Total}$ and Ge related constraints calculated for five independent configurations at 1050 K for the 22% system.

**Supplementary note 4: Validation of the dynamic behavior**

In order to focus on the transport properties, one computes first the mean-square displacement of a tagged atom of type $\alpha$ in the Ge-Se liquids:

$$\langle r^2(t) \rangle = \frac{1}{N_\alpha} \sum_{i=1}^{N_\alpha} \langle |r_i(t) - r_i(0)|^2 \rangle$$

(Supplementary equation 1)

with $N_\alpha$ is the number of atoms α. In a log-log plot of the mean square displacement with time, the diffusive regime can be detected from a linear behaviour with a slope of one in the long time limit, and such a regime is detected for the 1050 K liquid in GeSe$_2$ for t > 5ps. The diffusion constant D can be obtained via the Einstein relation limit from D=<r$^2$(t)>/6t.

We are not aware of any diffusivity data for Ge-Se melts. However, in order to check for the accuracy of our models, we can estimate an approximate diffusion constant from viscosity data using the Stokes-Einstein relation η=k$_B$T/6πrD where 2r=ρ$^{-1/3}$ involves the liquid densities ρ [6]. Note that a similar relationship (Eyring), D=k$_B$T/λη, where λ is a typical hopping length for the diffusing atom can be used [7]. In silicate liquids e.g., Na$_2$SiO$_3$, this Eyring equation holds very well with λ=2.8 Å, a distance typical of Si-Si and O-O separation in these melts, and this conclusion has also been demonstrated from a separate estimate of diffusivities and Green-Kubo viscosities using molecular simulations [8].

From experimental viscosity data of Ge-Se melts, one is able to proceed in a similar way and deduce such approximate diffusivities from the Stokes-Einstein equation, i.e. we obtain a direct comparison between our simulated diffusivities and experimental viscosity determined diffusivities. For instance, for the temperature 1373 K, a diffusion constant equal to $4.70 \times 10^{-5}$ $cm^2.s^{-1}$ can be calculated for $GeSe_2$, which compares favorably (same order of magnitude) to the calculated one from viscosity data (~0.002 Pa.s) of Stolen (~1300 K, [9]) which leads to $D_{exp}=3.2 \times 10^{-5}$ $cm^2.s^{-1}$.

**Supplementary note 5: Relaxation behavior**

The relaxation behaviour of the Ge-Se liquids has been investigated by analyzing the time dependence of the intermediate scattering function $F_s(k,t)$ at a wavevector (2.10 Å$^{-1}$) corresponding to the main peak of the static structure factor S(k) (Supplementary Figure 2). This function follows the Fourier components of density correlations and can be investigated for different conditions (Supplementary Figures 4 and 5). For a fixed composition and at low temperature (T<800 K), $F_s(k,t)$ exhibits the usual relaxation behavior with a β-relaxation plateau (e.g. at $F_s(k,t)$~0.2 for T=800 K in $Ge_{20}Se_{80}$, Supplementary Figure 4), followed by an α-relaxation behavior, detectable only for T>800K. This trend is similar to many other glass-forming systems including silica [10].

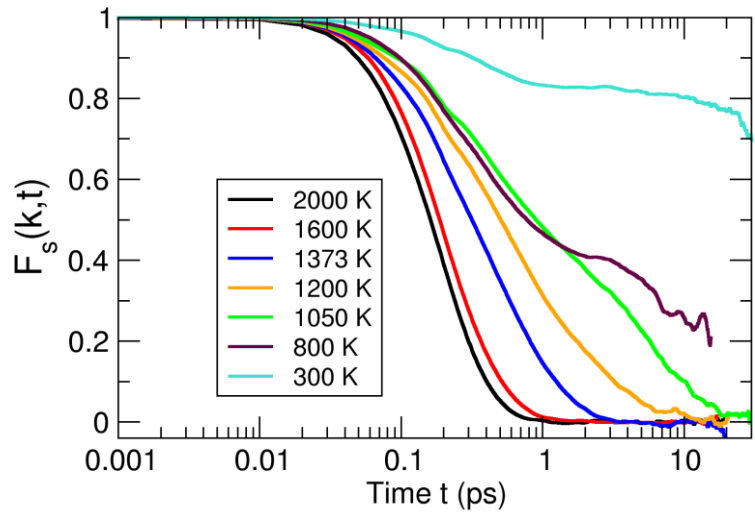

***Supplementary Figure 4: Time dependence of $F_s(k,t)$ of $Ge_{20}Se_{80}$ at different temperatures showing the single step relaxation process at high temperature (2000 K), and the onset of a two-step relaxation process for T<1050 K. For the present purpose, the isotherm 1050 K is the lowest temperature that leads to a diffusive regime, and that can be used for fitting the α-relaxation to obtain the relaxation time $\tau_\alpha$ (Fig. 1b).***

The same relaxation behavior can be investigated as a function of composition along an isotherm (1050 K, Supplementary Figure 5), and it is seen that composition does not affect the time evolution of the different intermediate scattering functions in a monotonic fashion since e.g. $Ge_{22}Se_{78}$ displays the slowest time evolution for $F_s(k,t)$ exhibiting even a starting β-relaxation plateau (e.g. at $F_s(k,t)$~0.5 in Supplementary Figure 5), followed by an α-relaxation behavior, also detectable at long times for all compositions. In the α-relaxation regime, (typically when $F_s(k,t)<e^{-1}$), the behavior of $F_s(k,t)$ can be fitted with a Kohlrausch stretched exponential function ($exp[-t/\tau_\alpha)^\beta]$, broken line) to yield the structural relaxation time $\tau_\alpha$ with composition (**Fig. 1d**, right axis).

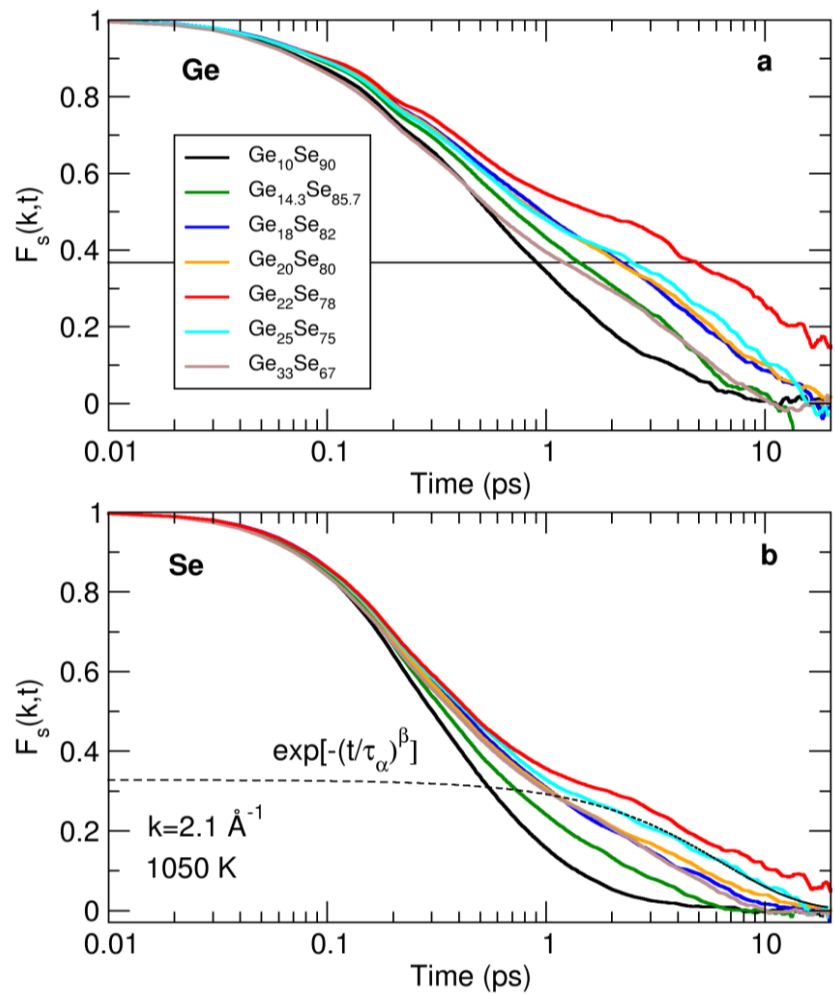

*Supplementary Figure 4: Time dependence of $F_s(k,t)$ for select different compositions. The thin solid line represents 1/e, and Kohlrausch exponential fits (broken line) have been performed in the long time limit, i.e. when $F_s(k,t)<1/e$.*

**Supplementary references**